\documentclass[journal,12pt,onecolumn,draftclsnofoot,]{IEEEtran}
\usepackage[T1]{fontenc}
\usepackage{verbatim}
\usepackage{graphicx}
\usepackage{threeparttable}
\usepackage{float}
\usepackage{amsmath}

\ifCLASSINFOpdf
\else
\fi
\usepackage{amsmath}
\usepackage[cmintegrals]{newtxmath}
\begin{document}
\bibliographystyle{unsrt}
%
\title{Deep Neural Network Aided Scenario Identification in Wireless Multi-path Fading Channels}
%
%
%

\author{Jun Liu,~Kai Mei,~Dongtang Ma,~Jibo Wei~\\
School of Electronic Science\\
National University of Defense Technology\\
Changsha,~China\\
Email: liujun15@nudt.edu.cn



}

%
%

\markboth{IEEE~Wireless~Communications~Letters,~Vol.~*, No.~*, *~2018}%
{Shell \MakeLowercase{\textit{et al.}}: Bare Demo of IEEEtran.cls for IEEE Communications Society Journals}
%



\maketitle

\begin{abstract}
This letter illustrates our preliminary works in deep nerual network (DNN) for wireless communication scenario identification in wireless multi-path fading channels. In this letter, six kinds of channel scenarios referring to COST 207 channel model have been performed. 100\% identification accuracy has been observed given signal-to-noise (SNR) over 20dB whereas a 88.4\% average accuracy has been obtained where SNR ranged from 0dB to 40dB. The proposed method has tested under fast time-varying conditions, which were similar with real world wireless multi-path fading channels, enabling it to work feasibly in practical scenario identification.
\end{abstract}

\begin{IEEEkeywords}
Deep neural network, scenario identification, wireless multi-path fading channels.
\end{IEEEkeywords}

%
\IEEEpeerreviewmaketitle

\section{Introduction}
%
%
%
%

\IEEEPARstart{T}{raditionally}, designing a radio receiver on the basis of theoretical analysis includes structuring its hardware, demodulation algorithm and parameters. Therefore, these invariable receivers may not perform satisfactorily in real commuication scenarios, which differ from channel models. Hopefully, software defined radio (SDR) devices has been increasingly applied in various domains like commerce and military because of its flexible configuration. In this light, it is feasible to extract characteristic parameters from wireless channels and then adjust the architecture of radio devices to fit real communication scenarios. Nevertheless, it is extremely difficult to estimate channel parameters and to identify communication scenario in real world, especially in time-varying wireless multi-path channels.


During the past twenty years, methods for scenario identification in time-varying channels have been proposed by serval researchers. Specifically, wireless channels can be intricately described by deterministic parameters like number of multi-path, path delays and delay spread. Additionally, statistic parameters like probability density functions (PDFs) and autocorrelation functions (ACFs) of fading envelopes are used to characterize channels' statistical property. Given relationship between these parameters and channel scenarios is non-linear, it is complicated to establish the mapping relationship. To identify channel scenarios, DSP-based methods which involve two steps is utilized in traditional setting. Extraction of key feature parameters like number of multi-path, path delays and delay spread is done, followed by minimization of characteristic vector distance algorhtim or Back-Propagation (BP) neural network \cite{1}. However, these methods cannot distinguish different scenarios which have similar or even same deterministic feature parameters but different statistic characters. 

Deep neural network (DNN) is an important research point in machine learning, showing excellent performance in various domains, especially in feature identification \cite{2}. Therefore, if channels with same statistic parameters cannot be distinguished by DSP-based methods, training DNN by channels statistic parameters could be considered as an alternative.

This letter proposes a DNN aided algorithm to identify wireless communication scenario in wireless multipath fading channels. Section I illustrates the necessity and difficulty of scenario identification in wireless multipath fading channels. Section II derives the system model. Section III gives the numerical simulation results. Section IV concludes the paper.

\section{SYSTEM MODEL}
The system model is divided into four sections, pre-processing, procedure A, procedure B and procedure C (Fig. 1) due to necessity of estimating channel parameters before identifying channel scenario.

\begin{figure}[H]  
  \centering  
  \includegraphics[height=4.0cm]{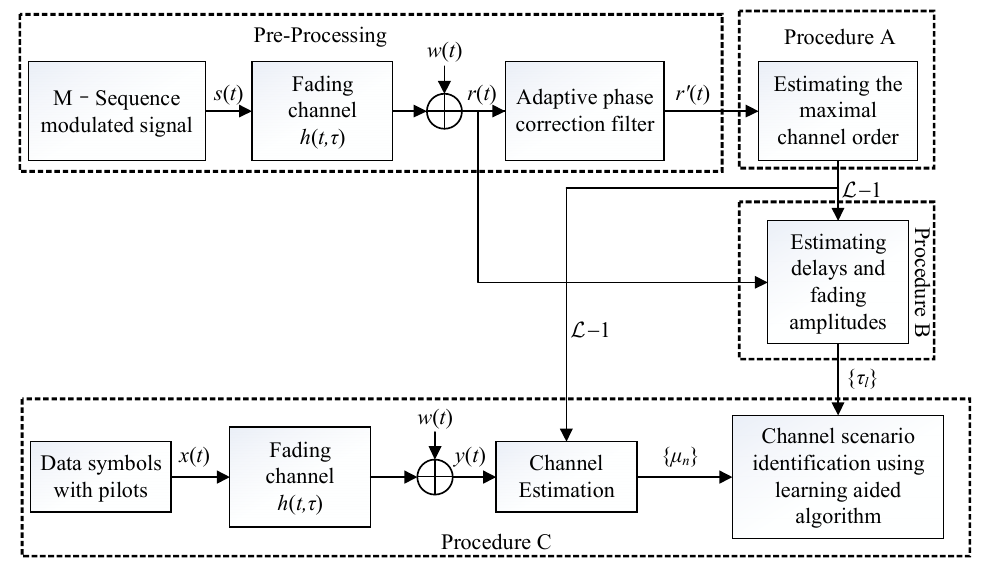}\\  
  \caption{System model for channel parameters estimation and scenario identification.}  
  \label{fig:1}  
\end{figure}  

As shown in (1),
\begin{equation}
h\left( {t,\tau } \right) = \sum\limits_{l = 0}^{{\cal L} - 1} {{\mu _l}\left( t \right)\delta \left( {{\tau} - {\tau _l}} \right)} 
\end{equation}
time-varying wireless multi-path fading channels are determinedly described by channel impulse response (CIR) $h(t,\tau )$.  ${\mu _l}\left( t \right)$ and ${\tau _l}$  represent fading amplitude and delay of \emph{l}th path respectively. 

In terms of channel scenario identification, there are massive algorithms assuming that the maximal channel order ${\cal L}$ and path delays ${\tau _l}$ for receiver are known. However, these parameters are usually unknown in real scenarios. Therefore, it is necessary to estimate ${\cal L}$ and ${\tau _l}$ before estimating ${\mu _l}\left(t \right)$. In this system model, the ${\cal L}$ will be probed in pre-processing and procedure A, followed by ${\tau _l}$ being estimated in procedure B. Next come delay profile ${\cal L}$ and ${\tau _l}$ acquired in procedure C and then data symbols with pilots transmitted by system begin to estimate ${\mu _l}\left(t \right)$. Finally, channel scenario will be identified. Details of the above-mentioned procedures will be illustrated in the following sub sections correspondingly, namely A, B and C.

\subsection{Pre-Processing and Estimating the Maximal Channel Order}

Assume that different fading paths can be distinguished(${{\tau _{l + 1}}-{\tau _l}>}$symbol period). To estimate the maximal channel order, m-sequence transmission has been done due to the special property of its autocorrelation function. Assume that $s\left( t \right)$ is
\begin{equation}
s\left( t \right) = m\left( t \right)\cos \left( {\omega t + \theta } \right)
\end{equation}
where $m\left( t \right)$ is m-sequence with values of ${\rm{ + 1}}$ or $ - 1$  and $\cos \left( {\omega t + \theta } \right)$ is carrier signal. After being delayed and faded by channel, the receiver signal can be given as
\begin{equation}
\begin{split}
r\left( t \right)&= g\left( t \right) + w\left( t \right)\\
&= \sum\limits_{l = 0}^{{\cal L} - 1} {{\mu _l}\left( t \right)} m\left( {t - {\tau _l}} \right)\cos \left[ {{\omega _l}\left( {t - {\tau _l}} \right) + {\varphi _l}} \right] + w\left( t \right)
\end{split}
\end{equation}
Equation (3) shows that the received signal is the sum of delayed and amplitude faded signal of $s\left( t \right)$ where ${\omega _l}$, ${\varphi _l}$ and $w\left( t \right)$ are the medium frequency, random phase of $l$th path and additive white Gaussian noise (AWGN) respectively (in Fig. 1, the process of pulse shaping and carrier recovery are omitted).

In order to mitigate the impact of random phase ${\varphi _l}$, $r\left( t \right)$ should be disposed by adaptive phase correction as shown in Fig. 2
\begin{figure}[H]  
  \centering  
  \includegraphics[height=2cm]{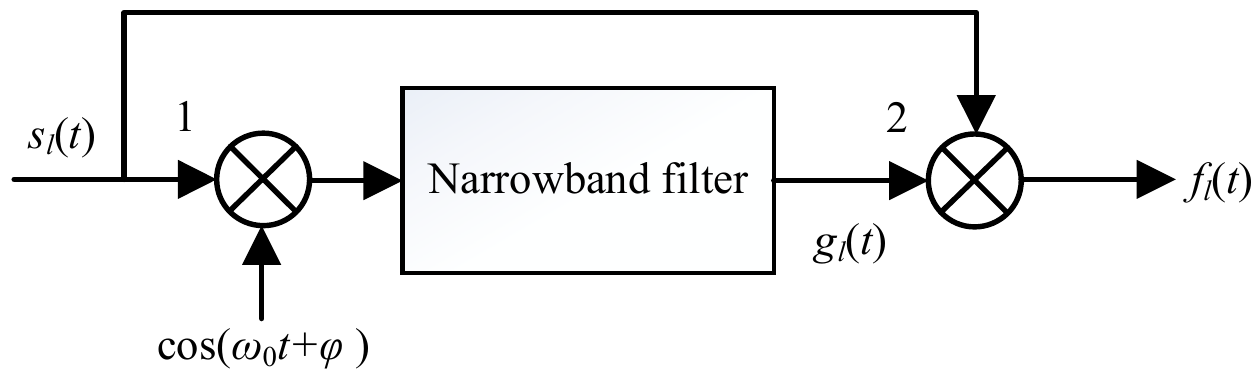}\\  
  \caption{Adaptive phase correction filter.}  
  \label{fig:2}  
\end{figure}  
where ${s_l}\left( t \right)$ is the $l$th delayed signal of $s\left( t \right)$. After being multiplied by unit 1 and filtered by narrowband filter sequentially, ${g_l}\left( t \right)$ can be given as
\begin{equation}
{g_l}\left( t \right) = {\mu _l}\left( t \right)\cos \left[ {\left( {{\omega _l} - {\omega _0}} \right)t - {\tau _l}{\omega _l} + {\varphi _l} - \varphi } \right].
\end{equation}
Only the beat frequency item is reserved after multiplying unit 2 (in Fig. 2, the band-pass filter for beat frequency is omitted), ${f_l}\left( t \right)$ can be expressed by
\begin{equation}
{f_l}\left( t \right) = \mu _l^2\left( t \right)m\left( {t - {\tau _l}} \right)\cos \left( {{\omega _0}t + \varphi } \right).
\end{equation}
The output signal of adaptive phase correction $r'\left( t \right)$ can be expressed as:
\begin{equation}
r'\left( t \right){\rm{ = }}\sum\limits_{l = 0}^{{\cal L} - 1} {\mu _l^2\left( t \right)m\left( {t - {\tau _l}} \right)\cos \left( {{\omega _0}t + \varphi } \right) + w'\left( t \right)} 
\end{equation}
where ${w'\left( t \right)}$ is narrowband Gaussian noise.

Fig. 3 shows the process of estimating the maximal channel order and Fig. 4 describes the correlation peak spectra. These are summarized as following steps:

Step 1: Inputting $r'\left( t \right)$ to envelope detection to remove the cosine item ${\cos \left( {{\omega _0}t + \varphi } \right)}$ from $r'\left( t \right)$.

Step 2: Multiplying the output signal from envelope detection with local m-sequence $m\left( t \right)$.

Step 3: Counting the peaks of the multiplied signal and estimating the maximal channel order.
\begin{figure}[H]  
  \centering  
  \includegraphics[height=1.5cm]{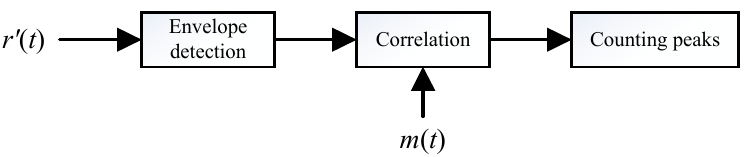}\\  
  \caption{Flow chart of estimating the maximal channel order.}  
  \label{fig:3}  
\end{figure}  
\begin{figure}[H]   
  \centering  
  \includegraphics[height=2cm]{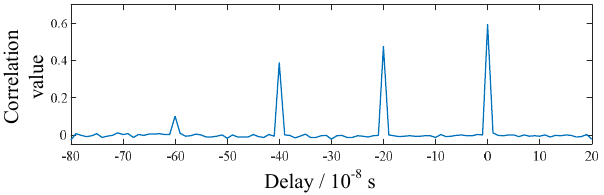}\\  
  \caption{Correlation peak spectra of a 4-delay-paths channel.}  
  \label{fig:4}  
\end{figure}  

\subsection{Estimating Delays and Fading Amplitudes}
In this subsection, system model in complex number field will be considered. Discretely, (3) can be rewritten as
\begin{equation}
r\left[ n \right] = \sum\limits_{l = 0}^{{\cal L} - 1} {{\mu _l}\left[ n \right]m\left[ {n - {\tau _l}} \right] + w\left[ n \right]} {\kern 1pt} {\kern 1pt} {\kern 1pt} {\kern 1pt} {\kern 1pt} {\kern 1pt} n = 1,2, \cdots ,N
\end{equation}
where $n$ represents the sample time at $n$Ts and Ts stands for the sample period given the assumption that ${\tau _l} = k{{\rm{T}}_{\rm{s}}}$ where $k$ is integer. Then, we supposed that fading amplitudes were complex values and did not change obviously in one delay estimation procedure. Therefore, ${\mu _l}\left[ n \right]$ could be written as ${\mu _l}$ while the influence of random phase is inflected by the phase of complex value ${\mu _l}$. In order to estimate delays and fading amplitudes more accurately, (7) can be converted into frequency domain:
\begin{equation}
R\left[ k \right] = \sum\limits_{l = 0}^{{\cal L} - 1} {{\mu _l}\exp \left( { - j{\omega _k}{\tau _l}} \right)M\left[ k \right] + W\left[ k \right]{\kern 1pt} {\kern 1pt} {\kern 1pt} {\kern 1pt} {\kern 1pt} {\kern 1pt} k = 1,2, \cdots ,N} 
\end{equation}
where $R\left[ k \right]$, ${M\left[ k \right]}$ and ${W\left[ k \right]}$ are the discrete Fourier transform of $r\left[ n \right]$, $m\left[ n \right]$ and $w\left[ n \right]$ respectively. Let ${{\rm{C}}_1}\left( {\left\{ {{\mu _l},{\tau _l}} \right\}_{l = 0}^{{\cal L} - 1}} \right)$ be the non-linear minimum mean-square cost function 
\begin{small}
\begin{equation}
{{\boldsymbol{C}}_1}\left( {\left\{ {{\mu _l},{\tau _l}} \right\}_{l = 0}^{L - 1}} \right){\rm{ = }}\sum\limits_{k =  - N/2}^{N/2 - 1} {{{\left| {R\left[ k \right] - M\left[ k \right]\sum\limits_{l = 0}^{L - 1} {{\mu _l}\exp \left( { - j\frac{{2\pi {\tau _l}k}}{N}} \right)} } \right|}^2}} .
\end{equation}
\end{small}
Let
\begin{equation}
{\bf{R}} = {\left[ {R\left( { - N/2} \right),R\left( { - N/2 + 1} \right), \cdots ,R\left( {N/2 - 1} \right)} \right]^T}
\end{equation}

\begin{equation}
{\bf{M}} = {\rm{diag}}\left\{ {M\left( { - N/2} \right),M\left( { - N/2 + 1} \right), \cdots ,M\left( {N/2 - 1} \right)} \right\}
\end{equation}

\begin{small}
\begin{equation}
{\bf{\alpha }}\left( {{\tau _l}} \right) = \left\{ {\exp \left[ { - j\frac{{2\pi {\tau _l}}}{N}\left( { - N/2} \right)} \right], \cdots ,} \right.\left. {\exp \left[ { - j\frac{{2\pi {\tau _l}}}{N}\left( {N/2 - 1} \right)} \right]} \right\}
\end{equation}
\end{small}
Equation (9) can be rewritten as matrix format:
\begin{equation}
{{\boldsymbol{C}}_2}\left( {{\mu _l},{\tau _l}} \right) = \left\| {{{\bf{R}}_l} - {\mu _l}{\bf{M}\boldsymbol{\alpha }}\left( {{\tau _l}} \right)} \right\|
\end{equation}
where
\begin{equation}
{{\bf{R}}_l} = {\bf{R}} - \sum\limits_{i = 1,i \ne l}^{{\cal L} - 1} {{{\hat \mu }_i}} {\bf{M}\boldsymbol{\alpha }}\left( {{{\hat \tau }_l}} \right)
\end{equation}
According to (13), ${{{\hat \mu }_l}}$ and ${{{\hat \tau }_l}}$ could be estimated by iterative algorithm:
\begin{equation}
{{\hat \mu }_l} = {\left. {\frac{{{{\boldsymbol{\alpha }}^H}\left( {{\tau _l}} \right)\left( {{{\bf{M}}^*}{{\bf{R}}_l}} \right)}}{{\left\| {\bf{M}} \right\|_F^2}}} \right|_{{\tau _l} = {{\hat \tau }_l}}}
\end{equation}

\begin{equation}
{{\hat \tau }_l} = \mathop {\arg \max }\limits_{{\tau _l}} {\left| {{{\boldsymbol{\alpha }}^H}\left( {{\tau _l}} \right)\left( {{{\bf{M}}^*}{{\bf{R}}_l}} \right)} \right|^2}.
\end{equation}
The iterative algorithm can be summarized as following steps:

Step 1: Assuming ${\cal L}=1$ and then estimating value ${{\hat \mu }_1}$ and ${{\hat \tau }_1}$ according to (15) and (16).

Step 2: Assuming ${\cal L}=2$ and estimating value ${{\hat \mu }_2}$ and ${{\hat \tau }_2}$ by utilizing ${{\hat \mu }_1}$ and ${{\hat \tau }_1}$ in Step 1 and then update the ${{\hat \mu }_1}$ and ${{\hat \tau }_1}$ by utilizing ${{\hat \mu }_2}$ and ${{\hat \tau }_2}$ . Iterating this calculation until results become convergent.

Step $l$: $ \cdots $

Step ${\cal L}$: Estimating ${\mu _{\cal L}}$ and ${\tau _{\cal L}}$ and then iterating the above steps until convergence appears \cite{3}.

In addition, the set of delay $\left\{ {{\tau _{\cal L}}} \right\}$ can be seen as a constant in a given scenario. Whereas, $\left\{ {{\mu _{\cal L}}} \right\}$ will not be considered as fixed value unless being performed in one specific estimation procedure. Consequently, the iterative algorithm is mainly used for delays estimation while channel estimation method is utilized in $\left\{ {{\mu _{\cal L}}} \right\}$ estimation due to cost-effectiveness.

\subsection{Deep Neural Network Aided Channel Scenario Identification}
Fig. 5 shows the stochastic characters of channels. To be specific, a CIR sample of COST207 4-paths Rural Area (RA) fading channel in delay and time dimension is illustrated (Fig. 5 (a)). Although two scenarios share the same delay characters (the maximal channel order and delays), they may differ from each other in statistic characters within each path, leading to difficulties of channel scenario identification. The Fig. 5 (b) shows the PDFs of channel in each path, which are complex to estimate in practical communication. To overcome the obstacle, this paper proposed using delay-discrete probability distribution plots (D-DPDPs) to reflect characters of channels, including delay and statistic character (Fig. 6). The most crucial advantage of D-DPDPs is that it could give statistical information about delays and envelopes of each fading path of channel. Values of each block represent probabilities of envelopes in a specific interval. For instance, the first row in Fig. 6 means the corresponding discrete probability distribution of fading envelopes of the channel which have a $0{\rm{\mu s}}$ delay path. Additionally, sum of values within each row was 1. It could be seen from Fig. 6 (a) and Fig. 6 (b) that the two D-DPDPs have different patterns, indicating that D-DPDPs could identify different statistic characters of fading envelope in the same scale of delay characters. In this light, channel estimation seemed quite essential. A basis expansion model-least square (BEM-LS) algorithm \cite{4,5}, rather than parameter estimation method in Procedure B, has been used to estimate channel, owing to the latter one is mainly used to estimate delays and may not give accurate CIR estimation here.

\begin{figure}[H]   
  \centering  
  \includegraphics[height=5.0cm]{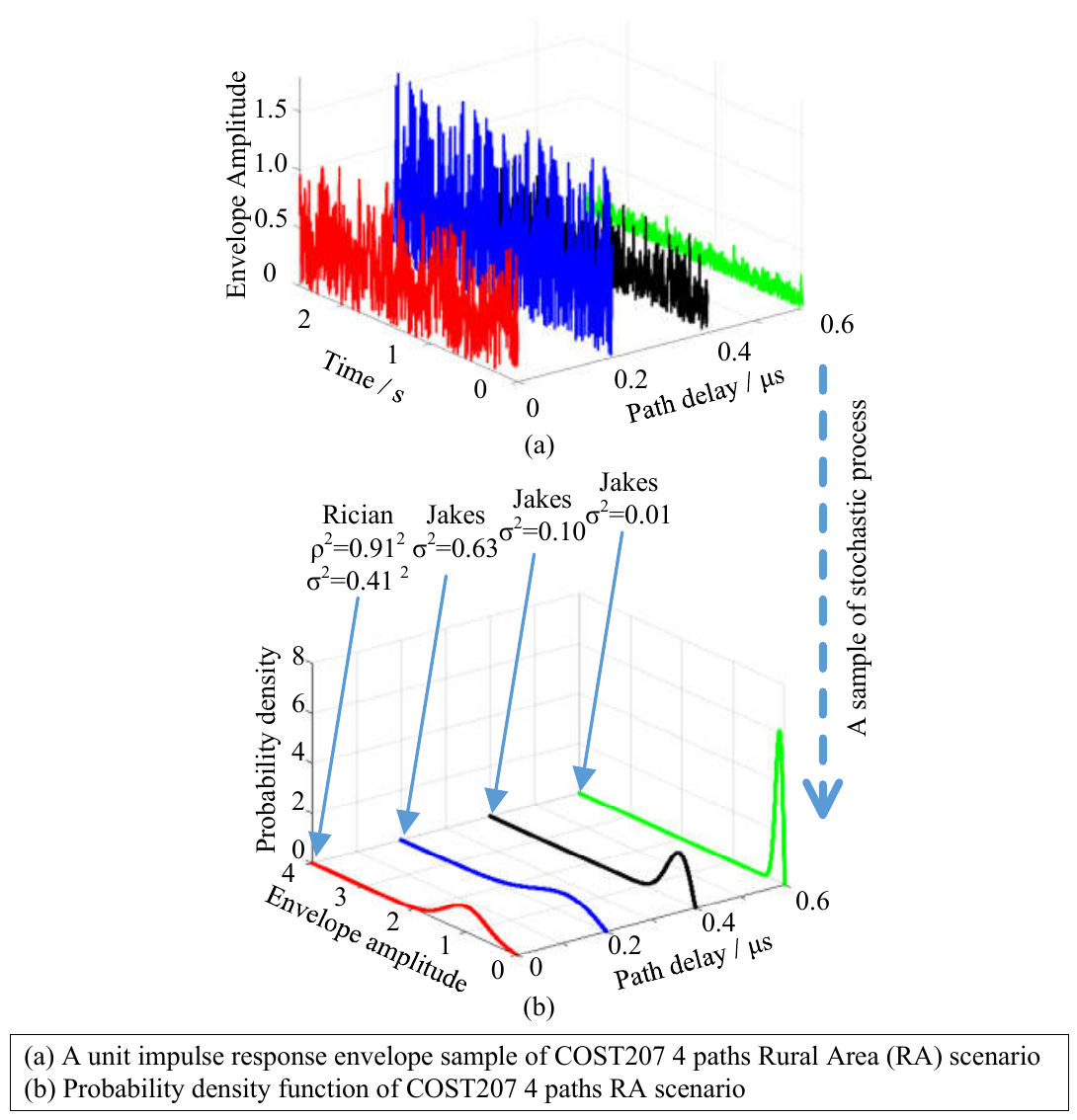}\\  
  \caption{Relationship between PDFs of the channel and a unit impulse response sample of the channel.}  
  \label{fig:5}  
\end{figure}  

\begin{figure}[H]   
  \centering  
  \includegraphics[height=4.0cm]{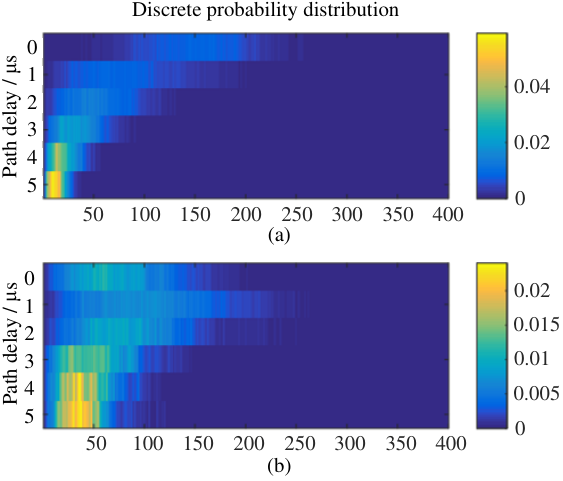}\\  
  \caption{Two D-DPDPs corresponding to two different 6-paths scenarios (one interval represents 0.05 in this figure).}  
  \label{fig:6}  
\end{figure}  

In order to identify channel scenario by DNN, it is necessary to convert D-DPDPs into vector format. Fig. 7 shows the configuration of DNN. Each D-DPDP in DNN data set is expressed as a one-dimensional vector ${\bf{i}}$ of 400${\cal L}$ (${\cal L}$ delay units in maximum~$\times $~400 probability intervals ranging from 0 to 2) bincounts by concatenating all of its rows. Similarly, for each D-DPDP, an 6$\times $1 binary one-hot vector ${\bf{o}}$ (with single non-zero element whose location indicates the channel scenario pertaining to that D-DPDP) is obtained.

\begin{figure}[H]  
  \centering  
  \includegraphics[width = 8.0cm]{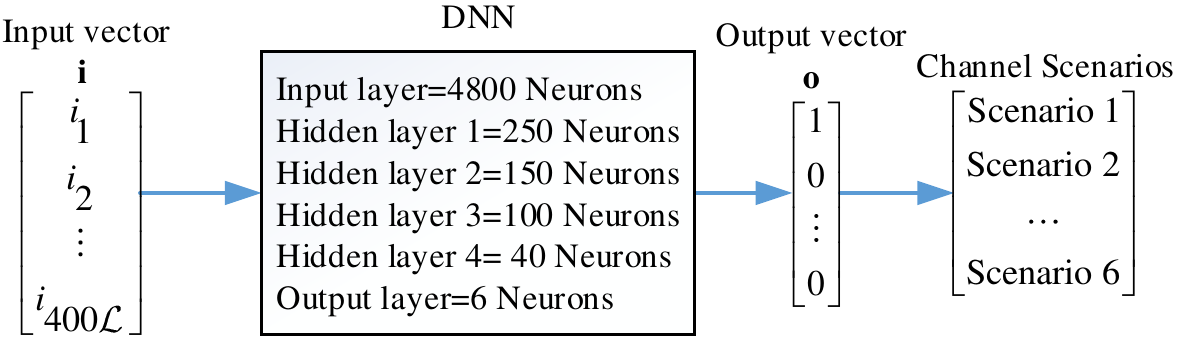}\\  
  \caption{DNN with bin-count vector ${\bf{i}}$ as input and channel scenario identification as output.}  
  \label{fig:7}  
\end{figure}  



\section{NUMERICAL RESULTS}
The system model utilized for numerical simulations has been shown in Fig. 1. There are mainly 3 steps in Procedure C. To begin with, transmitter will send data symbols with pilots by QPSK modulation mode. While CIR is estimated by BEM-LS algorithm, discrete prolate spheroidal sequences \cite{Zemen2005Time} were used as basis function. Six channel scenarios have been chosen for evaluation (TSBLE I). In particular, only average path gain and Doppler spectrum of COST 207 channel models have been reffered. Apart from this, the delay of $l$th path was set to be 10$l$ ${\rm{\mu s}}$, which means that one delay unit represents 10 ${\rm{\mu s}}$.
\begin{table}[H]
\centering
\caption{CHANNEL MODEL LABEL AND PROFILE}
\label{table:1}
\begin{tabular}{llp{3.5cm}l}
\hline
Location & Channel model label & Profile                                  \\ \hline
1        & cost207RAx4         & Rural Area (RAx), 4 taps                 \\
2        & cost207RAx6         & Rural Area (RAx), 6 taps                 \\
3        & cost207TUx6         & Typical Urban (TUx), 6 taps              \\
4        & cost207BUx6         & Bad Urban (BUx), 6 taps                  \\
5        & cost207HTx6         & Hilly Terrain (HTx), 6 taps              \\
6        & cost207BUx12     & Bad Urban (BUx), 12 taps                 \\ \hline

\end{tabular}
\end{table}
Except for delays and scenarios with same delays but different Doppler spectrum, all scenarios in TABLE I only reffered to the average path gains and Doppler spectrums of COST 207 channel models.
\begin{figure}[H]  
  \centering  
  \includegraphics[height=2.0cm]{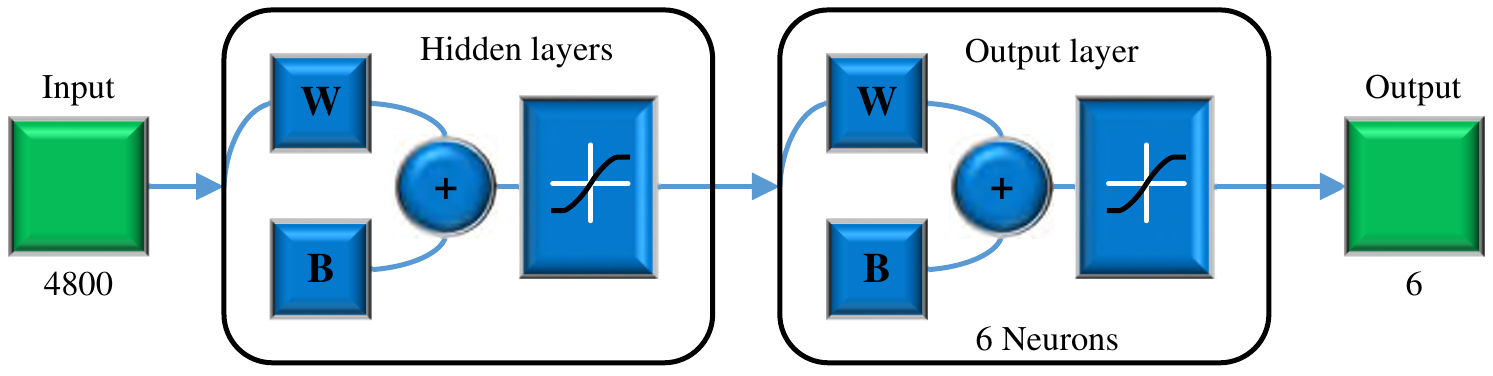}\\  
  \caption{Block diagram of DNN.}  
  \label{fig:8}  
\end{figure}  
Configuration of the DNN is shown in Fig. 8 where the blocks with letter ``W'' and ``B'' represent sets of weight and bias respectively. Four hidden layers and one output layer have been transferred by Tansig function in the DNN. A vector set of 4800 D-DPDPs, corresponding to 6 channel scenarios (100 vectors per scenario at 5 SNRs range from 0dB to 40dB and without noise) has been obtained (by symbol rate at 0.1M/s and the normalized Doppler frequency at 0.004). Each vector has been obtained from 25600 points unit impulse response estimated values of channels. Six hundred vectors without noise have been used to train DNN while other vectors corresponding to 5 SNRs have been used to test the DNN. If a vector obtained from the scenario with location lable 1 was identified as scenario with location lable except 1, it represents an error identification. The identification results for these test cases are summarized in TABLE II.
\begin{table}[H]
\centering
\caption{IDENTIFICATION ACCURACIES FOR VARIOUS SNRS}
\label{table:2}
\begin{tabular}{lllllll}
\hline
SNR / dB      & 0     & 10    & 20    & 30    & 40    & Avg.  \\ \hline
Accuracy / \% & 58.7 & 83.3 & 100 & 100 & 100 & 88.4 \\ \hline
\end{tabular}
\end{table}
Over 90\% identification accuracy in three out of five SNRs has been achieved by the proposed algorithm (TABLE II). When SNR was 10dB, the corresponding accuracy was 83.3\% which seemed tolerably well. The mean classification accuracy of 5 SNRs in this work was 88.4\%.

To conclude, compared with traditional channel scenario identification methods, the proposed algorithm can perform better accuracies under real world propagation conditions, where the signals are impaired by both AWGN and fast time-varying fading. In addition, its time complexity is 
\begin{equation}
\sum\limits_{h = 1}^{H - 1} {O\left( {2M{q_h}{q_{h + 1}}} \right)}  + O\left( {2M{q_H}{q_o}} \right) + O\left( {2M{q_o}} \right)
\end{equation}
where $H$ and $M$ stand for the number of hidden layer and training sample respectively, ${{q_h}}$ and ${{q_o}}$ represent the number of neuron in $h$th hidden layer and output layer respectively. Cost-effectiveness and simplicity have also been improved by the proposed algorithm since good performance could be achieved by using little points estimated values of channels.


\section{CONCLUSION}
We illustrated a DNN-based machine learning aided scenario identification algorithm in wireless multipath fading channels. The results indicated that this method could give accurate identification in several scenarios, which exist in real world. The proposed algorithm has been found workable in distinguishing channels with same deterministic feature parameters but different statistic characters. Moreover, cost-effectiveness could be improved and procedures could be simplified given this algorithm being implemented in software defined radio systems.



%




\ifCLASSOPTIONcaptionsoff
  \newpage
\fi

\end{document}